\documentclass[modern]{aastex631}
\usepackage{xspace}

\newcommand\teff{\ensuremath{T_\mathrm{eff}}\xspace}
\newcommand\logg{\ensuremath{\log g}\xspace}
\newcommand\msun{\ensuremath{M_\odot}\xspace}

\newcommand\avgamp{\ensuremath{\left < \, \mathrm{A} \, \right >}\xspace}
\newcommand\sdssj{SDSS J1557+0411}
\newcommand\wdpg{PG 1312+099}

\shorttitle{}
\shortauthors{Williams et al.}
\accepted{2025 October 20}
\submitjournal{the AAS Journals}

\graphicspath{{./}{figures/}}

\begin{document}

\title{84 Second and 169 Second Rotation of Two Isolated, Ultramassive, Strongly Magnetic White Dwarfs}

\author[0000-0002-1413-7679]{Kurtis A.\ Williams}
\affiliation{Department of Physics \& Astronomy \\
East Texas A\&M University  \\
P.O. Box 3011 \\
Commerce, TX 75429-3011, USA}

\author[0009-0000-2997-7630]{Zorayda Martinez}
\altaffiliation{Department of Astronomy, University of Texas at Austin}
\affiliation{Department of Physics \& Astronomy \\
East Texas A\&M University  \\
P.O. Box 3011 \\
Commerce, TX 75429-3011, USA}

\author{Melissa Ornelas}
\altaffiliation{REU participant at East Texas A\&M University}
\affiliation{Butler Community College\\ 
Andover Campus \\
715 E 13th St.\\
Andover, KS 67002, USA}


\begin{abstract}
Photometric variability in massive, magnetic white dwarfs (WDs) on the timescales of less than a few hours is oft interpreted as being due to magnetic spots on the surface of a rotating star.  Increasingly, numbers of these short period variables are being discovered with the continued growth of time-domain astronomy, testing theories of magnetic white dwarf formation and angular momentum evolution.  We present the detection of extremely rapid rotation in the WDs \sdssj~ and \wdpg, with periods of 168.94 s and 83.72 s, respectively.  We consider other possible causes of the monoperiodic photometric variability, including binarity (eclipses and ellipsoidal variations) and asteroseismic pulsations. Though these cannot be ruled out with the existing data, these alternative explanations seem unlikely.  \sdssj~ was predicted to be a rapidly rotating oxygen-neon core WD by \citet{2022MNRAS.516L...1C}, a prediction borne out at least in part by our observations.  \wdpg~ was previously observed via polarimetry to be rotating with a period of $\sim 5.4$ h; we propose that this object may be an unresolved double degenerate system. 
\end{abstract}

\section{Introduction} \label{sec.intro}

Time-series photometry of seemingly isolated, strongly magnetic, ultra-massive white dwarfs (WDs) has revealed an increasing number exhibiting short-period variability interpreted as rapid rotation with periods $P\lesssim 6\,\mathrm{h}$ \citep[e.g.,][]{2015ApJ...814L..31K,2017ApJ...841L...2H,2020ApJ...894...19R,2020MNRAS.499L..21P,2021ApJ...923L...6K,2021Natur.595...39C,2022AJ....164..131W}.  These discoveries have led to an emerging consensus that at least some strongly magnetic WDs are descended from and have their magnetic fields generated by mergers in double degenerate systems \citep[e.g.][]{2000PASP..112..873W,2008MNRAS.387..897T,2011PNAS..108.3135N,2012ApJ...749...25G}, though evidence suggests additional mechanisms such as WD core crystallization dynamos are also at work in some strongly magnetic WDs \citep[e.g.][]{2021NatAs...5..648S,2022MNRAS.516L...1C,2024MNRAS.528.6056H}.

As part of the myriad of investigations into isolated rotating magnetic WDs, we have targeted a large magnitude-limited sample of massive, magnetic WDs from the Montreal White Dwarf Database \citep[MWDD,][]{2017ASPC..509....3D} to search for rapid variability; sample details and initial results of this study will be presented in a forthcoming paper.  During the course of this study, we identified periodic, rapid ($\leq 3\,\mathrm{min}$) photometric variability in two WDs: \object{SDSS J155708.02+041156.4} (hereafter \sdssj) and \object{PG 1312+099} (Gaia DR3 3732584969752521984).  If we interpret the variability as rotation, these two WDs currently represent the second- and third-fastest rotating isolated WDs known; we present initial results, analyses, and interpretations in this paper.   

\section{Observations and Data Reduction}

We obtained time-series photometry of \sdssj~ on UT 2022 Jul 31 and UT 2022 Aug 1 with the ProEm frame transfer camera on the Otto Struve 2.1 m telescope at McDonald Observatory. On the night of 2022 Aug 1, exposures were obtained through alternating $g$ and $r$ filters with a 2 s dead time between exposures.  All other data consist of consecutive exposures through a BG40 filter with no dead time between exposures. 

Observations of \wdpg~ were obtained on UT 2023 Mar 27 followed by a shorter confirmation series obtained UT 2023 Jun 30.  All exposures were also obtained on the Struve telescope with ProEM and the BG40 filter. Observing logs are included in Table \ref{tab.obsdata}. 

On UT 2022 Jul 31, KAW left a faint dome light on for the first 2000 s of the series, resulting in higher than expected background counts and flat field artifacts being present in those images; we do not include these contaminated data in our further analysis.   We ensured the dome light was off for all subsequent observations.  

\begin{deluxetable}{llcccDDDD}
    \tablewidth{0pt}
    \tablecaption{Observing Log and Photometric Variability Parameters \label{tab.obsdata}}
    \tablehead{\colhead{UT Date} & \colhead{Target} & \colhead{Filter}  & \colhead{Exp.\ time} & \colhead{$N_\mathrm{exp}$} & \multicolumn{4}{c}{Frequency} & \multicolumn{4}{c}{Amplitude}  \\
     & & & (s) & & \multicolumn{2}{c}{$f$ (d$^{-1}$)} & \multicolumn{2}{c}{$df$ (d$^{-1}$)} & \multicolumn{2}{c}{$A$ (\%)} & \multicolumn{2}{c}{$dA$ (\%)}}
    \decimals
    \startdata
    2022 Jul 31 & SDSS J1557+0411 & BG40 & 10 & 1041 & 511.42 & 0.04 & 0.71 & 0.07 \\
    2022 Aug 01 & SDSS J1557+0411 & $g$ & 15 & 265 & 511.42 & 0.04 & 1.1 & 0.2 \\
    2022 Aug 01 & SDSS J1557+0411 & $r$  & 15 & 265 & 511.42 & 0.04 & 0.6 & 0.2 \\
    2023 Mar 27 & WD 1312+098 & BG40 & 10 & 2051 & 1032.0 & 0.2 & 0.7 & 0.1 \\
    2023 Jun 30 & WD 1312+098 & BG40 & 15 & 732 & 1032.7 & 0.8 & 0.8 & 0.1 \\
    \enddata
\end{deluxetable}

Data reduction and Fourier analysis for all observations are similar to those described in \citet{2022AJ....164..131W}.  In summary, all images were reduced and had aperture photometry performed using AstroImageJ \citep{2017AJ....153...77C}.  The aperture size was permitted to vary with seeing such that the aperture radius was $1.4\times \mathrm{FWHM}$.  All observation times were converted to Barycentric Dynamical Time.  Frequency analysis was performed using Period04 \citep{2005CoAst.146...53L}.  The detection threshold for potential signals was set to four times the average amplitude on the periodogram \citep[e.g.,][]{1993A&A...271..482B}.

\begin{figure}
    \centering
    \includegraphics[clip, trim=0cm 0cm 0cm 0cm, width=\textwidth]{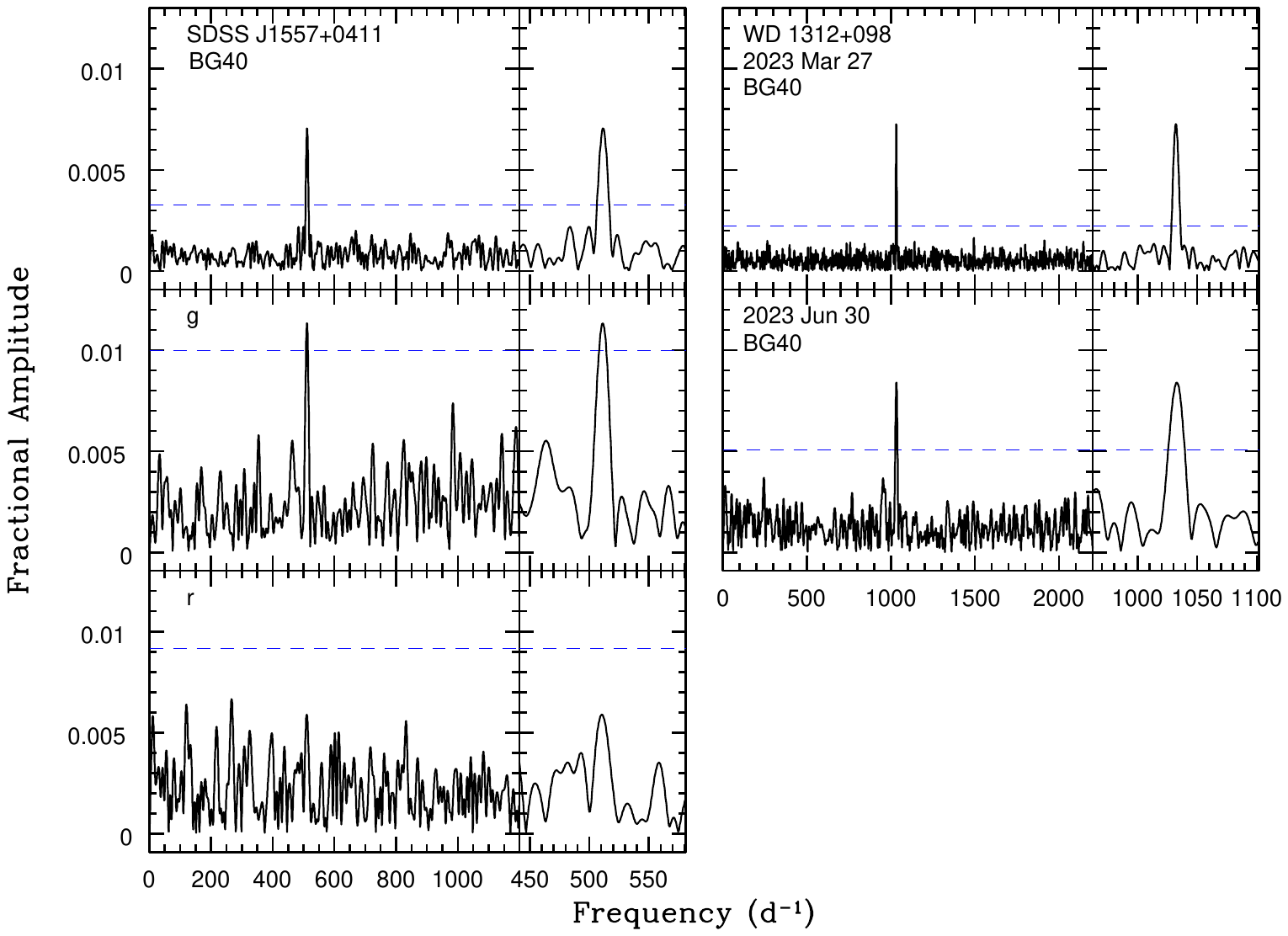}
    \caption{Periodograms for time series photometry of \sdssj~ (left panels) and \wdpg~ (right panels).  For each photometric series, an enlargement of the periodogram centered on the detected frequency is also included. A clear monoperiodic signal is detected well above the significance threshold in all cases except the $r$-band imaging of \sdssj.}
    \label{fig.dfts}
\end{figure}

The resulting periodograms are shown in Figure \ref{fig.dfts}.  For \sdssj, a significant signal is observed near a frequency of $\approx 511$ d$^{-1}$ in both the BG40 and SDSS $g$ periodograms, while for \wdpg~ a candidate signal is observed with a frequency of $\approx 1032$ d$^{-1}$.

For \sdssj, we determined a precise period for the modulation using a bootstrapping method similar to that used in \citet{2022AJ....164..131W}.  We thereby combined all photometric data from all filters to identify the single self-consistent alias at a frequency of $f=511.420\pm 0.035$ day$^{-1}$, corresponding to a period of $P=168.941\pm 0.011$ s.  We then refit each individual light curve keeping the frequency fixed in order to determine variability amplitudes in each filter.  These values are given in Table \ref{tab.obsdata}. Phases (not reported in Table \ref{tab.obsdata}) are consistent but not tightly constrained.  

We note that the $r$ band periodogram has no peak meeting the 4\avgamp critierion for finding previously unknown periods of variability. By fixing the period at 168.941 s, we are able to measure a photometric modulation in $r$ of $0.58\pm 0.17\%$, or $\approx 3.0\sigma$.  

\begin{figure}
    \centering
    \includegraphics[clip, trim=0cm 10.5cm 0cm 0cm, width=\textwidth]{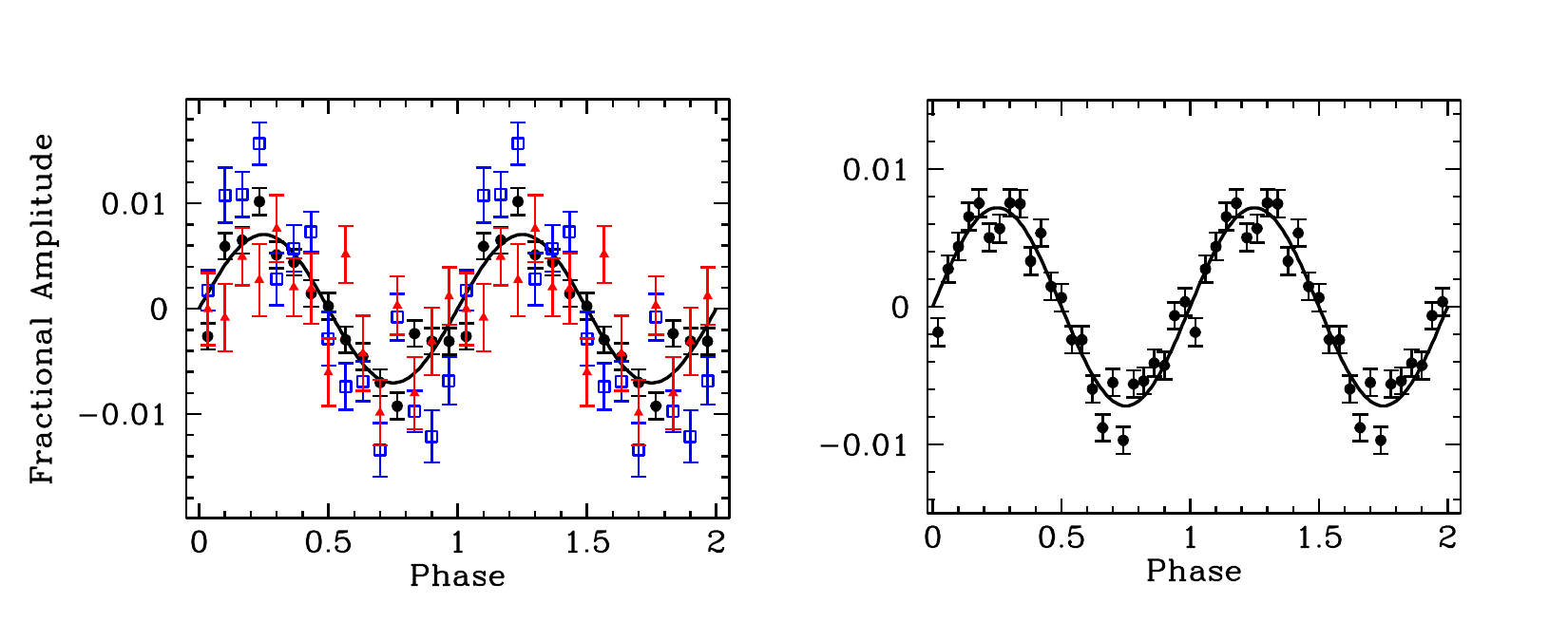}
    \caption{Pulse shapes for \sdssj~ (left) and WD 1312+098 (right) created by folding observed lightcurves on the primary frequency of modulation. The black curve in each panel is a sinusoid with the BG40 fit parameters from Table~\ref{tab.obsdata}.  Filled circles with error bars are points are BG40 data, open blue squares with error bars represent $g$ band data, and red filled triangles are $r$ band observations.  For \sdssj, the larger $g$ band amplitude is apparent.}
    \label{fig.pulse}
\end{figure}

The $r$-band data have a lower signal-to-noise due to the WD's spectral energy distribution. Additionally, it is observed in many rotating magnetic WDs that longer wavelength bandpasses exhibit smaller amplitudes of variability \citep[e.g.,][]{2022AJ....164..131W}.  The folded light curves in each filter are shown in Figure \ref{fig.pulse}.  

For \wdpg, the two nights of data are separated by three months, so no attempt is made to combine the data. The detected frequency from UT 2023 Mar 27 is $f=1032.0\pm 0.2\,\mathrm{d}^{-1}$, which corresponds to a period of $P=83.72\pm 0.02\,\mathrm{s}$, and a relative amplitude of $0.7\pm 0.1 \%$. Data from the shorter time series obtained in June exhibit a signal with a frequency and amplitude consistent with the March data.

\section{Discussion}
\subsection{Rapid rotation as the likely cause of observed variability}

We interpret the photometric modulation in both targets as being due to a spot on a rapidly-rotating WD with a rotation frequency equal to the photometric modulation frequency.  We admit the possibility that the true rotation period may be twice of the photometric period, if, for example, two spots are being detected, though the periodograms show no evidence for a signal at lower frequencies.  These rotation periods are among the fastest known for seemingly isolated WDs, with only the 70 s period of SDSS J221141.80+113604.4 known to be faster \citep{2021ApJ...923L...6K}.  At face value, this feeds into the generally accepted paradigm that strong magnetic fields in WDs are often generated from double degenerate mergers, at least in WDs hotter than the onset of core crystallization. 

Other mechanisms besides rotation could be responsible for rapid photometric variability; however, few if any of these seem plausible for the $\lesssim 3$ m periods observed here.   AM CVn systems have observed orbital periods as short as 5 minutes, though theoretically they could be as short as 100 s for double WDs \citep{2005ASPC..330...27N,2014LRR....17....3P}.  However, neither of our targets shows obvious accretion features in its optical spectrum, and neither is a known X-ray source. 

Detached WD-WD binaries are also known to exist with orbital periods as short as 12 minutes \citep[e.g.,][]{2011ApJ...737L..23B} and show photometric variation due to ellipsoidal variations, but such orbital periods are  significantly longer than the periods detected in \sdssj~ and \wdpg. While shorter periods for detatched binaries may be possible, the any such systems should come into contact on relatively small time scales due to significant gravitational wave radiation.  The possibility of detatched double degenerates could be tested either with high-cadence optical spectroscopy or future LISA observations.

Asteroseismic pulsations are another source of short-term photometric variability in WDs, with periods as short as $\approx 120\, \mathrm{s}$ for DA WDs on the blue edge of the instability strip, though shorter period pulsations may be possible in ultramassive WDs due to general relativistic effects \citep{2023MNRAS.524.5929C}.  However, both WDs lie outside the ZZ instability strip.  \sdssj~ has measured parameters of $\teff=24800\, \mathrm{K} \pm 1220\,\mathrm{K}$ \citep{2023MNRAS.520.6111H}. 

\wdpg's parameters have not, to our knowledge, been determined with spectroscopic models including magnetic effects. Results from non-magnetic, DA spectroscopic fits include: $\teff=22976$ K, $\logg=8.511$, $M=0.942\msun$ from \citet{2024A&A...682A...5V}; $\teff=20853$ K, $\logg=8.40$, $M=0.86\msun$ from \citet{2023A&A...677A.159T}, and $\teff=21292$ K, $\logg = 8.43$, $M=0.89\msun$ from \citet{2021MNRAS.508.3877G} -- all outside the instability strip.
We note, though, that these fits assume an isolated white dwarf at the distance given by the Gaia parallax. If
\wdpg~ is an unresolved binary (see Section \ref{sec.binary}), it remains possible that one component could be in the instability strip.  

\subsection{SDSS J1557+0411: a rapidly rotating O/Ne core WD with a magnetic field generated by crystallization?}

Six weeks prior to our observations of \sdssj, \citet{2022MNRAS.516L...1C} published a preprint discussing the possibility of determining the core composition of ultra-massive WDs via their magnetic fields; their preprint was unknown to us at the time of our observations.  In that paper, they predicted that \sdssj~ may a ``very fast" rotator.  

This prediction is based on the supposition that strong magnetic fields in ultra-massive WDs are caused by a dynamo in a crystallizing oxygen-neon (O/Ne) core. However, the magnetic field strength of \sdssj~ exceeds the predicted field strength based on an extrapolation of the dynamo scaling law of \citet{2009Natur.457..167C} by over a factor of 20. According to \citet{2022MNRAS.516L...1C}, such an exceedance could be explained if the Rossby number, $R_0$, $\ll 1$.  They estimate $R_0\sim {P}/{\tau_c}$, where $P$ is the rotational period and $\tau_c$ is the convective turnover timescale of $\sim 500\,\mathrm{s}$ for a crystallizing O/Ne core.  Based on our measurements of the rotational period, for \sdssj~ $R_0\sim 0.3$.  Therefore, the rapid rotation of \sdssj~ is consistent with the hypotheses of \citet{2022MNRAS.516L...1C}.

\subsection{The Conundrum of \wdpg \label{sec.binary}}

The 84 s period of variability in \wdpg~ is somewhat confounding, as \wdpg~ has a measured rotational period  from polarimetry of $5.428394\pm 0.000012\,\mathrm{h}$ at a semi-amplitude of 1.08\% \citep{1991ApJ...366..270S}. Within quoted error bars, the photometric frequency is inconsistent with being a high-order harmonic of the polarimetric frequency.

The photometric variability in \wdpg~ was detected significantly on two separate nights three months apart.  A search of telescope engineering data on both nights, including autoguiding measurements, stepper motor commands, and pointing data, found no signals consistent with the observed frequency of variability.  The ProEM camera uses a frame transfer CCD with no shutter, a frame transfer time of 148 ms, and GPS-triggered readout resulting in exposure timing precision of $\leq 100\,\mu\mathrm{s}$ \citep{2005JApA...26..321M}, so the measured variability frequency is inconsistent with being an alias of exposure spacing or detector properties.  We therefore are confident that the photometric variability is intrinsic to the star.

\citet{2023MNRAS.520.6111H} classify \wdpg~ in the class of shallow-line magnetic WDs and speculate that such WDs may have mixed atmospheres due to rapid rotation, or that they may be unresolved binaries. The optical spectrum of the object is complex and shows changes over the polarimetric rotation of the star suggesting a wide range of magnetic field strengths on the visible surface \citep{1997ApJS..112..527P}.  Clearly something is very odd about this star.  

If we assume that both the observed photometric and polarimetric periodicities are real, correct, and intrinsic to the star, we are led to the hypothesis that \wdpg~ may be a double degenerate system containing two magnetic, rapidly rotating magnetic WDs, one with the 84 s photometric period and one with the 5.4 h polarimetric period.  If all strongly magnetic WDs form from double degenerate mergers, this could suggest that the progenitor system was a hierarchical double binary system.  Still other explanations are plausible, though, including spin-up due to mass transfer during earlier stages of binary evolution and/or a He+CO WD merger in a triple system.

Given the complexity and variability of the optical spectrum and the wide range of binary and atmospheric parameters that could explain the relatively few observational constraints of \wdpg, it is not currently possible to be certain of the exact nature of the photometric variabilities, let alone have high confidence in a binary explanation of the system as a whole.  Depending on the orbital separation, future Gaia data may be able to detect orbital motion.  High-cadence spectroscopy and/or high cadence \mbox{(spectro-)polarimetry} may also be able to provide further insight into this unique object.

\section{Conclusions and Insights}

Searches for rapid photometric variability in magnetic white dwarfs have now led to the discovery of $\lesssim 3\,\mathrm{min}$ monoperiodic variability in three massive magnetic white dwarfs.  The two we present in this paper result from observations of just 25 massive magnetic white dwarfs, indicating that this rapid variability is not extraordinarily rare.  

Multiple mechanisms might produce such rapid variability, including nonadiabatic pulsations, and the existing data are not sufficient to prove that the variability is due to rapid rotation.  However, we feel the rotational interpretation of the variability to be the most plausible at present.  Additional observations, such as high speed, time-resolved polarimetry and spectroscopy, could provide data needed for confirmation of the root cause of variability.

The existence of very rapid rotating WDs such as \sdssj~ and \wdpg~ present stringent tests to hypotheses for the origin(s) of high magnetic fields in WDs, especially regarding the evolution of angular momentum. The unclear nature of \wdpg~ also points to potentially complex dynamical histories of magnetic WD progenitor systems.  Larger, well-defined samples of WD rotation sensitive to $\lesssim 60\,\mathrm{s}$ variability are required to ascertain the true population and parameter distributions of these extreme objects.

\begin{acknowledgements}
This work has made use of data from the European Space Agency (ESA) mission {\it Gaia} (\url{https://www.cosmos.esa.int/gaia}), processed by the {\it Gaia} Data Processing and Analysis Consortium (DPAC, \url{https://www.cosmos.esa.int/web/gaia/dpac/consortium}). Funding for the DPAC has been provided by national institutions, in particular the institutions participating in the {\it Gaia} Multilateral Agreement. We are very grateful to J.~Landstreet and S.~Bagnulo for discussions regarding \wdpg~ and to J. Kuehne for engineering data from the Struve telescope. The authors are grateful to the anonymous referee who provided useful suggestions, including some of the alternative explanations for the formation of \wdpg~ given in Section \ref{sec.binary}.

This material is based upon work supported by the National Science Foundation under Grant No.~ AST-1910551 and REU Grant No.~PHY-2050277. Any opinions, findings, and conclusions or recommendations expressed in this material are those of the authors and do not necessarily reflect the views of the National Science Foundation.

\end{acknowledgements}

\bibliography{170s_rotator}{}
\bibliographystyle{aasjournal}

%

\end{document}